# Chemical potential of quasi-equilibrium magnon gas driven by pure spin current

V. E. Demidov[1*], S. Urazhdin[2], B. Divinskiy[1], V. D. Bessonov[3], A. B. Rinkevich[3], V. V. Ustinov[3,4], and S. O. Demokritov[1,3]

[1]*Institute for Applied Physics and Center for Nonlinear Science, University of Muenster, Corrensstrasse 2-4, 48149 Muenster, Germany*

[2]*Department of Physics, Emory University, Atlanta, GA 30322, USA*

[3]*Institute of Metal Physics, Ural Division of RAS, Yekaterinburg 620041, Russia*

[4]*Institute of Natural Sciences, Ural Federal University, Ekaterinburg, 620083, Russia*

We show experimentally that the spin current generated by the spin Hall effect drives the magnon gas in a ferromagnet into a quasi-equilibrium state that can be described by the Bose-Einstein statistics. The magnon population function is characterized either by an increased effective chemical potential or by a reduced effective temperature, depending on the spin current polarization. In the former case, the chemical potential can closely approach, at large driving currents, the lowest-energy magnon state, indicating the possibility of spin current-driven Bose-Einstein condensation.

The discovery of the room-temperature magnon Bose-Einstein condensate (BEC) in magnetic insulators driven by parametric pumping [1] has spurred intense experimental and theoretical studies of this phenomenon [2-11]. It is now well established that the chemical potential of the magnon gas is increased by the parametric pumping, resulting in the formation of BEC when it reaches the lowest-energy magnon state [1]. The formation of magnon BEC has been experimentally confirmed by the observation of the spontaneous narrowing of the population function in the energy [3,11] and phase space [4]. Moreover, phase coherence of this state has been confirmed by the observation of interference in the real space [6].

While parametric driving provides a convenient approach to studies of BEC, it also has shortcomings. In particular, the energy of magnons injected by the parametric pumping is concentrated within a narrow interval [12], initially producing a strongly non-thermal state of the magnon gas. A significant thermalization time is required before a quasi-equilibrium state with non-zero chemical potential is formed [2,3]. The accompanying increase of the effective temperature of low-energy magnons can also be detrimental to the formation of BEC [7].

The magnon gas [13] can be also driven by the injection of spin current, for instance, generated by the spin-Hall effect (SHE) [14-16]. Spin current affects the magnon populations throughout the entire spectrum [13,17], which can avoid the non-thermalized transient states inherent to parametric driving. Moreover, the effects of spin currents on the magnon gas have been theoretically described [18-20], and experimentally interpreted [21] in terms of the variation of the chemical potential, suggesting the possibility of BEC formation at sufficiently large currents. While the spontaneous formation of coherent magnon state due to the spin current injection has been observed [22-28], there is no direct evidence for the quasi-equilibrium state that can be treated thermodynamically and form a BEC.

Here, we utilize a Permalloy/Pt bilayer to study the effect of pure spin current on the magnon distribution over a significant spectral range, allowing us to demonstrate that this distribution can be described by the Bose-Einstein statistics expected for the quasi-equilibrium state, and determine the current-dependent chemical potential and effective temperature. We show that, for one polarization of the spin current, the effective temperature of the magnon gas becomes significantly reduced, while the chemical potential stays almost constant. In contrast, for the opposite polarization, the effective temperature remains nearly unaffected, while the chemical potential linearly increases with current until it closely approaches the lowest-energy magnon state.

The studied system comprises a 2 μm wide and 5 nm thick Pt strip overlaid by a 1 μm wide and 10 nm thick ferromagnetic Permalloy (Py) strip [Fig.1(a)]. The independently measured saturation magnetization of Py is $4\pi M_0$=10.2 kG, and the measured resistivities are 275 and 325 nΩ·m for Pt and Py, respectively. The system is magnetized by a static magnetic field $H_0$ applied along the Py strip. For the studied 15 μm-long strip, the inhomogeneous dipolar field is negligible in the active device area. The electric current $I$ flowing in Pt is converted by SHE into a spin current $I_S$ injected into Py through the Py/Pt interface. The magnetic moment carried by the spin current is either parallel or antiparallel to the Py magnetization $M$, depending on the direction of current, [29] resulting in a decrease or an increase of the magnon population, respectively [13].

We study the magnon population by the micro-focus BLS technique [30]. We focus the single-frequency probing laser light with the wavelength of 532 nm onto the surface of the Py strip, and analyze the light inelastically scattered from magnons. The measured signal – the BLS intensity – is directly proportional to the spectral density of magnons $\rho(\nu)=D(\nu)n(\nu)$, where $\nu$ is the magnon frequency, $D(\nu)$ is the density of magnon states weighted by the wavevector-dependent measurement sensitivity, and $n(\nu)$ is the occupation function [1].

A representative BLS spectrum recorded at $H_0$=200 Oe and $I$=0 exhibits a peak with the highest intensity in the frequency range ν=4-5.5 GHz, and a shallow high-frequency tail extending to 9 GHz [Fig. 1(b)]. The origin of these spectral features is elucidated by the analysis of magnon dispersion in the Py strip [Fig. 1(c)], calculated using the approach described in Ref. [30]. The spectrum is quantized in the direction perpendicular to the Py strip, and is continuous in the longitudinal direction. The allowed transverse wavevector components are $k_y \approx \pi m/w$, where $w$ is the width of the Py strip and a positive integer $m$ is the mode index. Because of the dependence of the BLS sensitivity on the wavevector, the fundamental mode $m$=1 provides the largest contribution to the BLS spectrum, producing the peak at ν=4-5.5 GHz, as indicated in Fig. 1(b) by the shadowed area. Despite the reduced BLS sensitivity to the higher-order modes, it allows the measurement of the magnon population in a broad range of frequencies up to ν=9 GHz. Our measurements also provide an opportunity to distinguish between the quasi-uniform ferromagnetic resonance (FMR) mode with $k_x$=0 and the lowest-energy finite-wavevector magnon state at frequency $\nu_{min}$ [Fig. 1(c)], since their frequency separation of about 0.8 GHz is significantly larger than the resolution of the BLS. As we shall see below, this distinguishing feature allows us to determine which of these states is predominantly over-populated due to spin current injection, addressing a long-standing debate in the studies of spin current-induced effects [27].

To reduce the effects of Joule heating on the current-dependent BLS spectra, in our measurements the current was applied in 200 ns long pulses with the repetition period of 1 μs, and the BLS spectra were accumulated over the duration of the pulse. The effects of current are illustrated in Figs. 2(a) and (b) for $I$=-20 and 20 mA, respectively. At $I$<0, the BLS intensity decreases, and the spectrum shifts to higher frequencies. In contrast, at $I$>0 the BLS intensity strongly increases, while the spectrum shifts to lower frequencies.

Solid squares in Fig. 2(c) show the current dependence of the intensity integrated over the measured spectrum, which characterizes the total number of low-energy magnons accessible to BLS. This dependence is consistent with the expected reduction/increase of the magnon population by the appropriate polarization of spin current [13].

The dependence of the lowest magnon frequency $\nu_{min}$ on current [open squares in Fig. 2(c)] can be attributed to a combination of the Oersted field of the current and the variation of the effective magnetization of Py due to the effect of spin current on the magnon population, as well as Joule heating of the sample. The calculated contribution of the Oersted field is shown in Fig. 2(c) by the dashed curve [31]. The experimental data closely follow this dependence at $|I|$<10 mA, and deviate from it at larger current magnitudes. The deviation is larger at $I$>0 than at $I$<0. Since Joule heating does not depend on the sign of current, we conclude that the total magnon population that determines the effective magnetization is significantly affected by the spin current [13].

We now analyze the spectral distribution of spin current-driven magnon population, by comparing the zero-current BLS spectra with those obtained at finite currents. At $I$=0, the magnon gas in thermal equilibrium, with the temperature equal to the experimental temperature $T_0$=295 K and the chemical potential $\mu$=0 [1]. Correspondingly, the measured BLS intensity is proportional to the weighted spectral density of magnons $\rho_0(\nu)=D(\nu)n_0(\nu)$, where $n_0(\nu)$ is the Bose-Einstein distribution. In the limit $h\nu \ll k_BT_0$, the latter is well approximated by the Rayleigh-Jeans law $n_0(\nu)=k_BT_0/h\nu$, where $k_B$ is the Boltzmann constant and $h$ is the Planck constant. At finite spin current, $\rho_I(\nu)=D(\nu)n_I(\nu)$ with a current-dependent distribution $n_I(\nu)$. If the magnon gas is driven into a quasi-equilibrium state, this distribution can be written as $n_I(\nu)=k_BT_{eff}/(h\nu - \mu)$, with effective temperature $T_{eff}$ and chemical potential $\mu$. The density of states $D(\nu)$ is not expected to be influenced by the spin current, aside from the frequency shift discussed above. Therefore, the ratio of the BLS signals measured with and without current, or equivalently the frequency-dependent enhancement of the magnon population, is

$$R(\nu) = \frac{T_{\text{eff}}}{T_0} \frac{\nu}{\nu - \mu/h}. \tag{1}$$

This relation allows us to test whether the current-dependent magnon populations are well described by the quasi-equilibrium distribution, and extract the current-dependent values of $T_{\text{eff}}$ and $\mu$. Note that the roles of these parameters in Eq. (1) are qualitatively different: variations of $T_{\text{eff}}$ result in frequency-independent scaling of $R$, whereas $\mu>0$ produces a monotonically decreasing dependence $R(\nu)$.

Solid symbols in Figs. 3(a) and 3(b) show on the log-linear scale the BLS spectra recorded at $I$=-20 and 20 mA, respectively. Open symbols in the same plots show the spectrum obtained at $I$=0, shifted in frequency by the value determined from the data in Fig. 2(c). The data in Fig. 3(a) illustrate that at $I$<0, the magnon populations decrease approximately uniformly over the entire frequency range of the detected spectrum. In contrast, the increase of the population at $I$>0 [Fig. 3(b)] is more significant at lower frequencies, suggesting that the effects of opposite spin current polarizations are qualitatively different. Figure 3(c) shows the ratio of the spectra obtained with and without current. For $I$=-20 mA [open symbols in Fig. 3(c)], this ratio is independent of frequency. According to Eq. (1), this indicates that the dominant effect of spin current at $I$<0 is the reduction of the effective temperature, $T_{\text{eff}} \approx 0.76 T_0 = 224$ K at $I$=-20 mA. The frequency-dependent enhancement of the magnon population at $I$=20 mA [solid symbols in Fig. 3(c)] is also well described by Eq. (1). In this case, a good fit is achieved with $T_{\text{eff}} \approx T_0$, and the effective chemical potential in the frequency units $\mu/h$= 3.94 GHz [solid curve in Fig. 3(c)]. We note that the uncertainty of 20% for the effective temperature determined from fitting is significantly larger than the uncertainty of 5% for the chemical potential.

Figure 4 summarizes the results of the same analysis performed for different currents. The decrease of the effective temperature with increasing magnitude of $I<0$ (Fig. 4(a)) gradually saturates at large currents, which can be attributed to Joule heating that competes with the effects of spin current. Indeed, heat flow simulations show that the average increase of temperature in Py during the current pulse is about 45 K at $I=20$ mA, comparable to the temperature reduction induced by the spin current. At $I>0$, the effective chemical potential increases linearly up to $I=15$ mA [Fig. 4(b)], reaching 80% of $h\nu_{min}$ at this current. We note that the increase of temperature by 45 K at 20 mA provides a minor contribution to the magnon distribution, which is determined mostly by the increased chemical potential.

Extrapolating the linear dependence to larger currents, one would expect the chemical potential to reach the energy of the lowest magnon state at $I\approx17.5$ mA, which should result in the formation of the Bose-Einstein condensate of magnons. Instead, the growth of $\mu$ rapidly saturates at $I>15$ mA. This result is consistent with the previous studies, which showed that single-frequency current-driven magnetization dynamics cannot be achieved by injection of spin current into an extended region of the magnetic film [27], due to the onset of nonlinear magnon interactions in the strongly driven magnon gas that suppress the population of low-energy magnon states. It may be possible to overcome these effects by utilizing frequency-dependent magnon radiation losses [22], which are analogous to the evaporative cooling used in the experiments with atomic condensates [32].

Finally, we analyze the effects of the static magnetic field $H_0$ on the spin current-driven variations of the effective chemical potential. Measurements discussed above were performed at fields ranging from 100 to 500 Oe. While the observed behaviors remained similar over the entire field range, the efficiency of the chemical potential variation by the spin current strongly depended on field. Since the dependence $\mu(I)$ is linear at moderate $I>0$ [Fig. 4(b)], the spin-current efficiency can be characterized by the slope $d(\mu/h)/dI$, as shown by point-down triangles in Fig. 5. It rapidly increases with increasing small field, plateaus at $H_0$=300 Oe, and gradually decreases at larger fields. By extrapolating the linear dependence µ(*I*), we determine the critical electrical current density $J_C$ in Pt, at which the chemical potential would reach the energy of the lowest magnon state in the absence of the nonlinear suppression of magnon population [point-up triangles in Fig. 5]. This dependence exhibits a minimum at $H_0$=150 Oe, and a linear increase at larger fields. A similar dependence has been observed for the critical current in spin-Hall nano-oscillators [22,26]. We note that the latter are characterized by the auto-oscillation onset current densities that are very close to the values of $J_C$ extrapolated from our measurements. These similarities point to a close relation between the current-induced variation of the effective chemical potential and the current-induced auto-oscillations.

In conclusion, our experimental results provide a direct spectroscopic evidence that the magnon gas is driven by the pure spin current into a quasi-equilibrium state, which can be described by the Bose-Einstein distribution with current-dependent values of chemical potential and effective temperature. Our findings provide support for the theoretically proposed interpretation of current-induced magnetization auto-oscillations in terms of the Bose-Einstein condensation of magnons, which can be realized by avoiding the nonlinear magnon interactions that suppress the low-frequency mode populations at large magnon densities. Our results should stimulate further experimental and theoretical exploration of the relationship between the thermodynamics of magnon gases driven by spin currents and coherent magnetization dynamics.

This work was supported by the Deutsche Forschungsgemeinschaft, the NSF Grant Nos. ECCS-1509794 and DMR-1504449, and the program Megagrant № 14.Z50.31.0025 of the Russian Ministry of Education and Science.

**Figure captions**

FIG. 1 (a) Schematic of the experiment. (b) BLS spectrum of magnons in the Py strip measured at $I=0$ and $H_0=200$ Oe. Shadowed area shows the calculated BLS response for the fundamental magnon mode $m=1$. (c) Calculated dispersion spectrum of magnon modes in the Py strip. $m$ is the mode index, $\nu_{min}$ is the frequency of the lowest-energy magnon state.

FIG. 2 (a) and (b) Representative BLS spectra recorded at $I=-20$ and 20 mA, respectively, together with the reference spectrum obtained at $I=0$. (c) Current dependences of the BLS intensity integrated over the measured spectrum (solid squares) and of the frequency of the lowest-energy magnon state $\nu_{min}$ (open squares). Solid curve is a guide for the eye. Dashed line shows the calculated variation of $\nu_{min}$ due to the Oersted field of the current. The data were obtained at $H_0=200$ Oe.

FIG. 3 (a) and (b) BLS spectra recorded at $I=-20$ and 20 mA, respectively, (solid symbols) together with the reference spectrum obtained at $I=0$ (open symbols) shifted in frequency by the value determined from the data of Fig. 2(c). Note the logarithmic scale on the vertical axis. (c) The ratio of the spectra obtained with and without current. Dashed line is the mean value of the ratio for $I=-20$ mA. Solid curve is the fit of the ratio for $I=20$ mA by Eq. (3) with $T=T_0$ and $\mu/h=3.94$ GHz.

FIG. 4 (a) Current dependence of the effective temperature of the magnon gas at $I<0$. (b) Current dependence of $\mu/h$ (point-up triangles) and of the frequency of the lowest-energy magnon state $\nu_{min}$ (point-down triangles) at $I>0$. Curves are guides for the eye.

FIG. 5 Static-field dependences of $d(\mu/h)/dI$, the efficiency of spin current-driven chemical potential variation in frequency units (point-down triangles), and of $J_C$, the critical

current density in Pt at which the chemical potential is expected to reach the energy of the lowest magnon state (point-up triangles).

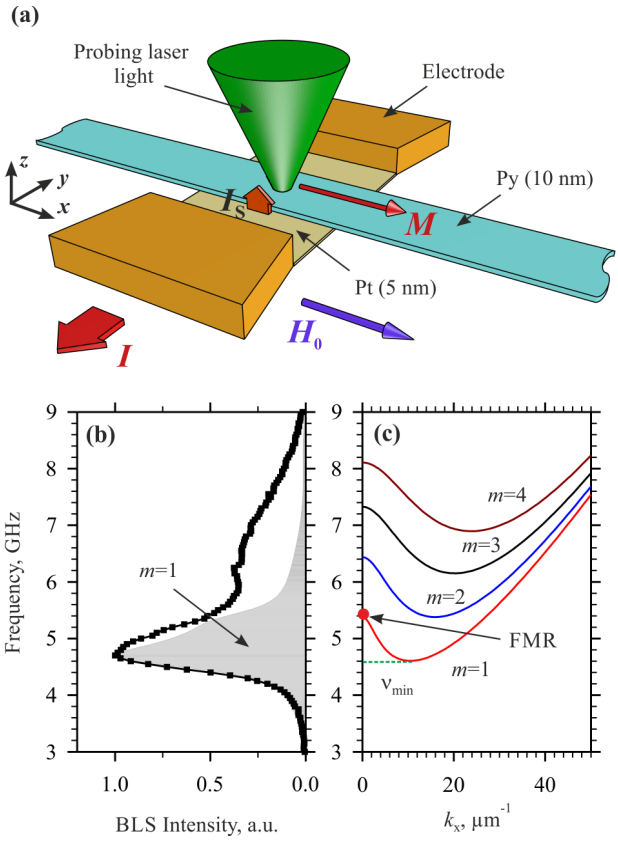

Fig. 1

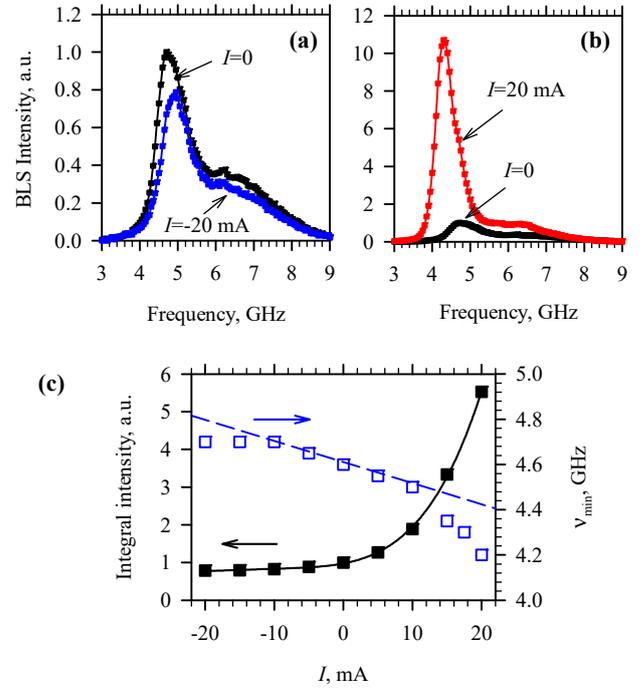

Fig. 2

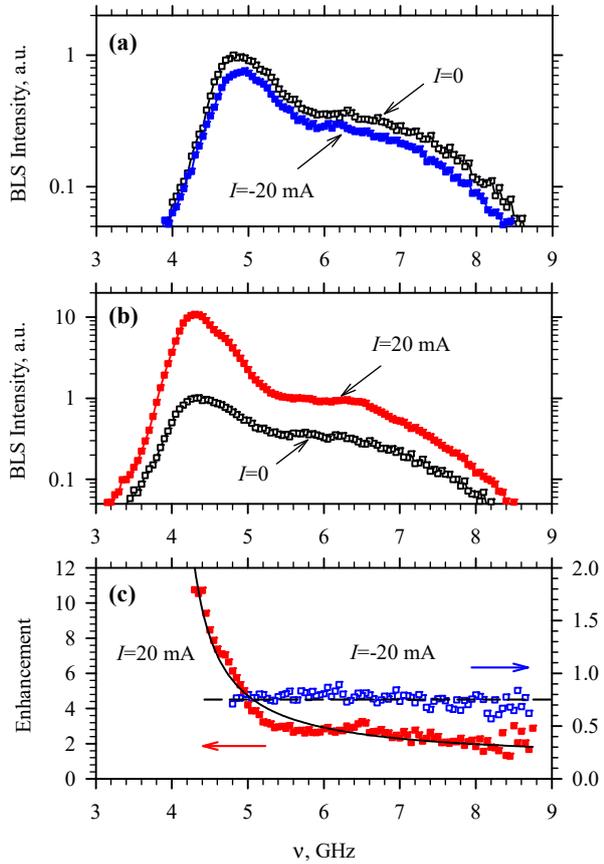

Fig. 3

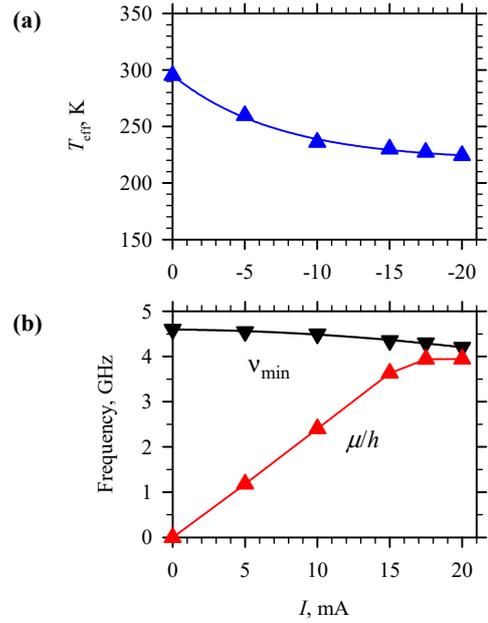

Fig. 4

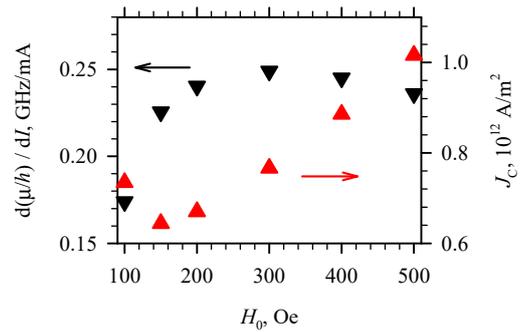

Fig. 5